\documentstyle[12pt,aaspp4]{article}

\lefthead{Bekki,  Beasley, Forbes, \& Couch}
\righthead{Formation of star clusters}

\received{2003 April 11}
\begin{document}
\title{Formation of star clusters in  
the LMC and SMC. I. Preliminary results on cluster formation from colliding gas clouds}

\author{Kenji Bekki} 
\affil{
School of Physics, University of New South Wales, Sydney 2052, Australia}

\author{Michael A. Beasley \& Duncan  A. Forbes} 
\affil{Centre for Astrophysics \& Supercomputing, Swinburne University
of Technology, Hawthorn, VIC,  3122, Australia}

\and

\author{Warrick J. Couch}
\affil{
School of Physics, University of New South Wales, Sydney 2052, Australia}

\begin{abstract}
We demonstrate that single and binary star clusters can be 
formed during cloud-cloud  collisions triggered by the 
tidal interaction between the Large and Small Magellanic clouds.
We run two different sets of self-consistent numerical
simulations which show that compact, bound star clusters can be formed
within the centers of two colliding clouds due to strong 
gaseous shocks, compression, and dissipation, providing the
clouds have moderately large relative velocities
($10-60$ km s$^{-1}$).
The impact parameter determines whether  the two colliding  clouds 
become  a single or a binary cluster.
The star formation efficiency in the colliding clouds is
dependent upon the  initial ratio of the relative velocity of the clouds to
the sound speed of the gas.
Based on these results, we discuss the observed larger  
fraction of binary clusters, and star clusters with high 
ellipticity, in the Magellanic clouds.
\end{abstract}

\keywords{galaxies: interaction --- 
galaxies: star clusters --- 
Magellanic Clouds
}

\section{Introduction}

It well established that several physical properties of
the globular clusters and populous young blue clusters in the Large Magellanic
Cloud (LMC) are differ markedly from those of clusters
in the Galaxy (e.g., van den Bergh 2000a).
These properties include the more flattened shapes of the LMC clusters (e.g., Geisler \& Hodge 1980;
van den Bergh \& Morbey 1984), the disky distribution of its globular cluster 
system (e.g., Schommer et al. 1992), a larger fraction of apparently binary clusters
or physical cluster pairs in the LMC
(Bhatia \& Hatzidimitriou 1988; Bhatia et al. 1991; Dieball \& Grebel 1998),
a possible  ``age/metallicity  gap''
(e.g., Da Costa 1991; Olszewski et al. 1991;
Geisler et al. 1997; Sarajedini 1998; Rich et al. 2001),
larger sizes at a given galactocentric distance (van den Bergh 2000b),
and the presence of a significant 
number of massive young to intermediate age clusters in the LMC (e.g., van den Bergh 1981, 2000a).

The higher fraction of binary clusters in the LMC, in particular,  has attracted
much attention from theoretical and numerical works.
Fujimoto \& Kumai (1997) proposed that oblique cloud-cloud collisions
in an interaction between the LMC and the Small Magellanic Cloud (hereafter, SMC)
can result in the formation of binary star clusters revolving around
each other.
Leon  et al. (1999) proposed a different scenario in which
the tidal capture of clusters in a group 
(where tidal encounters are  expected to be more common)  could be 
associated with the formation of the LMC binary clusters.
de Oliveira et al. (2000) suggested that merging of binary clusters
can be responsible for the observed flattened shapes of LMC clusters.

Kumai et al. (1993) and Fujimoto \& Kumai (1997) pointed out that 
if interstellar gas clouds are in large-scale disorganized motions 
with velocities of more than $50-100$ km s$^{-1}$ in the interacting LMC/SMC system,
then they may collide with one another  to form compact star clusters
due to strong shock compression. 
This idea is  supported  by
the coincidence between the observationally inferred two ``burst'' epochs
($\sim$ 100 Myr and $1-2$ Gyr ago 
of cluster formation 
after  the initial LMC collapse phase)
(e.g., Girardi et al. 1995)
and the theoretically predicted epochs of closest encounter between the LMC and  the SMC
(Murai \& Fujimoto 1980;  Gardiner et al. 1994; Gardiner \& Noguchi 1996). 
However, due to the lack of extensive numerical studies of this scenario,
these authors did not address  (1) whether the high-speed, oblique 
cloud-cloud collisions, which are central to the scenario of Kumai et al. (1993),
are present in the LMC/SMC interaction 
and (2) how star formation efficiency may increase in
the colliding clouds such that {\it compact and bound star clusters}
 rather than {\it unbound field  stars} will be formed.

In this paper, by using numerical simulations,
we demonstrate that the star formation efficiency of colliding 
gas clouds in interacting galaxies can significantly increase,
resulting in the formation of compact stellar systems.
This numerical investigation is two-fold:
We first derive  the most probable impact parameter and
the relative velocity of two colliding  clouds 
in a large-scale dynamical simulation of the interacting  LMC/SMC.
We then investigate the hydrodynamical evolution and star formation 
processes of  colliding clouds, based on the most probable
parameter values for cloud-cloud collision derived in our first set of simulations.
Here we describe the formation of star clusters in colliding
clouds in a general way, rather than attempt to 
explain precisely the observed physical properties of
star clusters for the LMC/SMC  system in a self-consistent manner. 
Since this paper is the first step toward the better
understanding of star cluster
formation in the LMC-SMC system,  the numerical models are somewhat idealized and
lack realism in some areas. In future papers, we will describe  the formation
processes, structure and kinematics, and chemical properties  of star clusters
formed from cloud-cloud collisions, using a more sophisticated simulation. 
The origin of disky distribution of LMC old star clusters will be 
also discussed in our future papers in terms of recent cosmological simulations
of globular clusters (e.g., Kravtsov \& Gnedin 2003).

\placefigure{fig-1}
\placefigure{fig-2}

\section{Models}

 We first describe the disk galaxy models in which the dynamical evolution
of gas clouds in interacting LMC/SMC is investigated. We
then describe the hydrodynamical models for
the evolution of colliding gas clouds with star formation 
in the LMC/SMC system.

\subsection{Tidal interaction in the LMC/SMC system}

We model LMC and SMC as bulge-less, gas-rich  disks
with an  initial gas mass fraction of 0.1. They Magellanic Clouds
are modeled in a fully self-consistent way by using
the Fall-Efstathiou (1980) model, with 
the exponential density profile for the disks
and halo dark matter-to-disk mass fraction equal to 4.
The total galactic mass and the disk size for the LMC (SMC) are assumed to be  
2.0 $\times$ $10^{10}$ $ \rm M_{\odot}$  
(2.0 $\times$ $10^{9}$ $ \rm M_{\odot}$)
and 7.5 kpc (2.4 kpc), respectively.
In order to investigate the nature of cloud-cloud collisions
in the interacting LMC/SMC,
we adopt the ``sticky particle method'' (e.g., Hausman \& Roberts 1984)
in which the interstellar medium is described as an ensemble of
discrete gas clouds.
The size ($r_{\rm cl}$) of an individual cloud with a given mass
($M_{\rm cl}$) is chosen such that the cloud satisfies the observed 
mass-size relation of gas clouds (Larson 1981).
All calculations related to  self-gravitating gas clouds
and stellar components 
were carried out on GRAPE systems (Sugimoto et al. 1990)
and total particle number in each simulation is 20000
for dark matter and  22000 for disk components.

We focus on the past 1 Gyr evolution of the LMC/SMC
(in particular, at the latest SMC's pericenter passage), during
which time populous young star clusters are known to have  formed (e.g., van den Bergh 2000a). 
Since the tidal effect on the LMC/SMC system due to the Galaxy is
small compared to that from the interaction 
between the Magellanic Clouds themselves at SMC's pericenter passage (Gardiner \& Noguchi 1996),
we do not explicitly include the gravitational effect of the Galactic
dark matter halo.
Guided by the earlier numerical results of
tidal interaction between the Galaxy, the LMC, and the SMC 
(Gardiner et al. 1994; Gardiner \& Noguchi 1996),
we choose a plausible set of orbital parameters for the   LMC/SMC.
The apocenter radius of the interaction  (for the last 1 Gyr)
is set to be 30 kpc. 
The pericenter of the orbit (represented by $R_{\rm p}$),  inclination of
the LMC's disk  with respect to the orbital plane (${\theta}_{\rm LMC}$),
and that of SMC (${\theta}_{\rm SMC}$) are assumed
to be free parameters. 
Although we investigated a number of models with different parameter values,
we present the results of the model
with $R_{\rm p}$ = 3.75 kpc, ${\theta}_{\rm LMC}$ = $-15^{\circ}$,
and ${\theta}_{\rm SMC}$ = $45^{\circ}$.
We choose this set of parameters since they exhibit 
behavior characteristic of cloud-cloud collisions (e.g., distribution of
relative velocity of clouds) in the present study.

The most important parameter of this model, involving interacting galaxies 
with a mass ratio of 0.1, is the pericenter of the SMC. If the pericenter
distance is large (e.g. 15 kpc, which is twice the LMC's disk
size), the frequency of cloud-cloud
collisions is  not enhanced  significantly in the simulation,  and the star formation rate
is not significantly increased in the simulated LMC disk.
Accordingly, the parameter of the pericenter should be carefully chosen.
We base the parameter values of the LMC/SMC orbital properties on
the early numerical simulations by Gardiner \& Noguchi (1996),
which are  not only consistent with the observed location and radial velocity
of the LMC/SMC system but also successful in reproducing the observed physical
properties of the Magellanic stream. 
The orbital parameters in
the present study (and thus in  Gardiner \& Noguchi 1996) are 
broadly consistent with $HIPPARCOS$ data by Kroupa \& Bastian (1997),
as shown by Yoshizawa \& Noguchi 2003.
Therefore, the adopted parameter values can be regarded as reasonable,
though the exact orbital properties of the LMC/SMC system have not been
observationally determined (thus the orbital parameters should be still free parameters). 
 
It is possible that introducing a mass spectrum on clouds instead of a single mass for all
clouds affects the results shown in this paper. The total number of cloud-cloud collisions
($N_{\rm cl}$) between the clouds with masses of $M_{\rm cl}$
 during the time interval $dt$ can be written as,
$N_{\rm cl}  = \pi  \times dt  \times {r_{\rm cl}}^2  \times V_{\rm cl} \times {\Phi}_{\rm cl}$,
where $r_{\rm cl}$, $V_{\rm cl}$, and $ {\Phi}_{\rm cl}$ are
the cloud radius, typical cloud velocity, and number density of gas clouds.
By assuming that the number density for a given volume is proportional to
the cloud number function observed in the Galaxy (e.g., Harris \& Pudritz 1994)
and adopting the Larson's mass-size relation (1981), 
we can derive the $M_{\rm cl}$ dependence of $N_{\rm cl}$.
Since $V_{\rm cl}$ is highly likely to be independent on $M_{\rm cl}$, 
the observed relations $ {\Phi}_{\rm cl} \sim {M_{\rm cl}}^{-1.63}$
and  $M_{\rm cl} \sim {r_{\rm cl}}^2$ imply that
$N_{\rm cl} \sim {M_{\rm cl}}^{-0.63}$. This derived relation
suggests that (1) cloud-cloud collision rates are larger for the clouds
with smaller masses and  (2) larger clouds can more frequently collide
with smaller clouds.  Therefore a spectrum of cloud masses is 
expected to lead to a larger number of low mass clusters.

\subsection{Star formation in colliding gas clouds}

Next we investigate the hydrodynamical evolution of two colliding clouds
by using a TREESPH code with star formation (Bekki 1997). 
The initial cloud mass ($M_{\rm cl}$) and  size ($r_{\rm cl}$)
are set to be $10^{6}$ $M_{\odot}$ and 97 pc, respectively,
which are consistent with the observed mass-size relation by Larson (1981),
and therefore with the large-scale simulations described in \S 2.
A gas cloud is assumed to have an isothermal radial density profile
with $\rho (r) \propto 1/(r+a)^2$, where $a$ is the core radius of the cloud
and set to be $0.2r_{\rm cl}$.
An isothermal equation of state with a sound speed of $c_{\rm s}$
is used for the gas, and $c_{\rm s}$ is set to 4 km s$^{-1}$
for models with $M_{\rm cl}$ = $10^{6}$ $M_{\odot}$.
We choose this value of $c_{\rm s}$ guided by the virial theorem
and Larson's mass-size relation. 
A gas particle in a given cloud is converted into a collisionless
stellar particle if two conditions are met: First, the local 
dynamical time scale (corresponding to  ${(4 \pi G\rho_{i})}^{-0.5}$,
where $G$ and $\rho_{i}$ are the gravitational constant 
and the density of the gas particle, respectively)
must be shorter than the sound crossing
time (corresponding to   $h_{i}/c_{\rm s}$, 
where $h_{i}$ is the smoothing length of the gas)
Secondly, the gas flow is converging.
This method therefore mimics star formation due to the 
Jeans instability in gas clouds.

The initial orbital plane of the two colliding clouds with relative
velocities of $V_{\rm r}$ and impact parameter of $b$
is set to be the $x$-$y$ plane. The position and the velocity of each  cloud
is represented by ${\vec{x}}_{i}$ 
and ${\vec{v}}_{i}$ ($i$ = 1, 2), respectively.
We generally only show our ``standard model'' in which
${\vec{x}}_{1}$ = ($-1.5r_{\rm cl}$,$-0.5b$,0) = $-{\vec{x}}_{2}$,
${\vec{v}}_{1}$ = ($0.5V_{r}$,0,0) = $-{\vec{v}}_{2}$,
$V_{\rm r}$  = 20 km s$^{-1}$ (or $V_{\rm r}/c_{\rm s}$ = 5), 
and $b$ = $0.5r_{\rm cl}$ (= 48.5 pc), 
since this model describes the typical behavior of star cluster
formation in colliding clouds in our simulations. 
Using the most probable values of $V_{\rm r}$ and $b$ derived from
our large-scale simulations, we investigate the parameter
dependencies of formation processes
of star clusters on  $V_{\rm r}$ and $b$ 
for 0 $\le$ $V_{\rm r}$  $\le$ 67 km s$^{-1}$
and 0 $\le$ $b/2r_{\rm cl}$  $\le$ 1. 
20000 SPH particles are used in a simulation.

Because our adopted total particle number is limited,
the resolution of the simulation is at most $10^2$ $M_{\odot}$ in mass
and $\sim$ 1 pc in scale for the models with $M_{\rm cl}$ = 
$10^6$ $M_{\odot}$. Therefore, a stellar particle converted from
a gas particle does not represent directly an individual star with a mass
and size the same as that observed.
Most of the stars in our simulations are formed in the very center
of a gas cloud, where the Jeans mass of the gas is of order $10^2-10^3$ $M_{\odot}$
due to the lower temperature and the higher gas density.
Therefore, the stellar particles in our simulations can be regarded as
small ``sub-clusters'' of stars with the mass of $10^2-10^3$ $M_{\odot}$  
Our future higher-resolution simulations with the total (gaseous) particle number
of more than $10^6$ will enable us to address not only the physical properties
of the sub-clusters, but also the formation of a single cluster from
the merging between these sub-clusters.

We choose an initial gas temperature (sound speed for isothermal gas)
guided by  the virial theorem for a gas cloud with a given mass and size.
In other words, the value of $c_{\rm s}$ is chosen 
such that an isolated gas cloud (not merging with another cloud)
is unable to collapse spontaneously. Accordingly, the isolated
gas model cannot
form stars in its central regions, and the fragmentation of such gas clouds
does not occur.
This ensures that if star formation occurs in colliding clouds,
it is purely a result of the hydrodynamical evolution of gas driven
by cloud-cloud collisions in our numerical study.
Thus, the adopted assumption of isothermal equation of state
and a higher initial gas temperature (corresponding to the virial temperature
of the gas) can help us to better interpret the derived results
of our numerical simulations. However, the reader should note
that this prescription is not as physically realistic as 
including heating and cooling sources in the gas.

\section{Results}

\subsection{The probability of high-speed, oblique cloud-cloud collisions}

Figure 1 shows how the frequency of cloud-cloud collisions can be enhanced
and, which type of cloud-cloud collisions most frequently occur, during
the LMC/SMC interaction.
As the SMC passes by the pericenter of the orbit 
at $T$ = 0.61 Gyr ({\it where T represents the time that has elapsed 
since the two disks begin to interact}), 
the strong tidal force induces non-axisymmetric 
structures (i.e., bars and spiral arms) in the disks of both
the LMC and SMC.
The large-scale tidal force randomizes the motion of the clouds 
during the interaction, and consequently the cloud-cloud
collision rate is increased  by a factor of $>$ 4 
after the pericenter passage. 
The cloud-cloud collisions with $V_{\rm r}$ 
$\simeq$ $25-40$ km s$^{-1}$
are  the most common  during the interaction (e.g., $T$ = 0.92 Gyr),
and we estimate the mean $V_{\rm r}$ to be 60 km s$^{-1}$.
The impact parameter represented by $b$ (Binney \& Tremaine 1987)
for cloud-cloud collisions can be widely distributed
during the tidal interaction, although collisions with 
larger values ($b/2r_{\rm cl}$ $>$ 0.5) represent $\sim$ 2/3
of the total.
These results demonstrate that high-speed ($V_{\rm r}$ $>$  50 km
s$^{-1}$), oblique collisions between two similar clouds are
likely in the LMC/SMC interaction, thus confirming the earlier
suggestions by Fujimoto \& Kumai (1998).

However, it should be stressed here  that
colliding gas clouds with relative velocities of more 
than $50-100$ km s$^{-1}$ are not particularly common 
amongst colliding gas clouds in our simulation.
Therefore,  the proposed large-scale motions with
velocity of more than $50-100$ km s$^{-1}$ (Kumai et al. 1993; 
Fujimoto \& Kumai 1997) are less likely in interacting LMC/SMC. 
The reason for the lower velocities here ($V_{\rm r}$ of $\simeq$ $25-40$
km s$^{-1}$, rather than $50-100$ km s$^{-1}$) 
is that the LMC's cloud system is not strongly disturbed
by the SMC tidal field due to the small mass ratio of the 
SMC to the LMC ($\sim$ 0.1).
Our numerical results shown in Figure 1 
suggest that if star clusters are formed from
cloud-cloud collisions in interacting LMC/SMC, 
then the clusters formed from clouds with 
relative velocities of more than $50-100$ km s$^{-1}$ constitute 
only a minor population amongst the ensemble of young clusters 
in the LMC/SMC systems.
The pros and cons of the original collisional formation
model of star clusters in the LMC/SMC system (Kumai et al. 1993)
are discussed in detail later.

\subsection{The formation of star clusters in colliding gas clouds}

Figures 2 and 3 illustrate how star clusters are formed
in two colliding clouds in our standard model.
Strong gas compression and dissipation during the collision leads
to an elongated slab-like structure formed at around  $T$ = 17.1 Myr,
where $T$ represents the time elapsed since the two
clouds began to collide.
As the dissipative merging proceeds,
the density of gas becomes very high in the shocked regions 
which are originally the central regions of the two clouds ($T$ = 22.8 Myr).
Two compact clusters are formed in these high-density gas regions
and begin to orbit each other  ($T$ = 22.8 Myr).
This result implies  that the orbital angular momentum
of the two gas clouds is efficiently converted into that of
the binary star clusters during the dissipative cloud-cloud collision.
The star formation is akin to an instantaneous ``starburst''
with a maximum star formation rate of 0.095 $M_{\odot}$ yr$^{-1}$, 
and 40 \% of the gas is converted into stars within 10 Myr.

A stellar particle is assumed to be  formed from each gas cloud
which is  considered to be collapsing  due to  gravitational  
instability (i.e., Jeans instability in the present study).  
The Jeans mass ($M_{\rm J}$) of gas in the central regions of colliding clouds
is estimated to be $\sim 10^3$ $M_{\odot}$, and, as such, each stellar
particle can be regarded as representing a small ``sub-cluster''.
Since all of these sub-clusters are formed in the very center of each
colliding cloud, the single massive cluster formed 
in each colliding cloud may be thought of as consisting 
of numerous small sub-clusters in the early
stages of massive cluster formation.
Unfortunately, because of the limited resolution of the present simulation, we
cannot investigate the subsequent dynamical evolution of these sub-clusters.
However, we may reasonably assume that these numerous sub-clusters
will finally merge with one another and consequently  erase
all of the substructures inside the cluster.
In fact, we do not observe any new stellar particles escaping
from the parent clouds because they are initially in the 
deepest potential well (i.e., the center of the clouds).
Thus a  single star cluster with a very smooth
and homogeneous mass distribution would finally form. 

The identification of such substructure in our simulations is
interesting when one considers recent evidence for the presence
subclustering in young cluster environments. Testi et al. (2000) 
identified three spatially and kinematically distinct subclusters within
Serpens, a nearby ($\sim$300pc) cloud comprising of 500-1500
M$\sun$ molecular gas and 40--80M$\sun$ young stellar objects
(Giovannetti et al. 1998). Our simulations support the idea that
the characteristic Jeans scale in cloud-cloud collisions gives
rise to several subclusters which then proceed to merge in a bottom-up
(hierarchical) fashion. The idea that star formation in stellar  clusters 
proceeds preferentially in "sub-clusters" of enhanced stellar
density was suggested by Clarke, Bonnell \& Hillenbrand (2000).

The formation process described in our simulations  differs 
in important ways from that proposed originally by Kumai et al. (1993).
Kumai et al. (1993) envisaged the formation of star and globular clusters 
from gravitationally unstable gas with $M_{\rm J}$  of $10^5-10^6$ $M_{\odot}$
in colliding clouds.
Our simulations suggest  that a single massive cluster is initially not
a single massive cluster but a cluster of numerous small ``sub-clusters''
that are formed from gas with smaller $M_{\rm J}$ 
(significantly smaller than $10^5-10^6$ $M_{\odot}$).
These clusters are born in the very centers of colliding  clouds,
and may finally form a single massive cluster.
Previous theoretical works argue that the very low  $M_{\rm J}$ 
(thus very small cluster mass) in the shocked gas layer is a
serious problem for the cloud-cloud collision model of star cluster formation
(e.g., Kumai et al. 1993).
The present numerical results suggest that this Jeans mass problem
may not be so important. The incipient sub-clusters 
may  quickly merge with one another to form a single massive cluster
due to their compact distribution in the central regions
of the colliding clouds.

We may speculate that the long-term evolution of binary clusters,
of which a detailed discussion is beyond the scope of this paper,
may depend on whether or not the remaining gas is quickly removed
during/after the cloud collisions. Such gas removal likely 
occurs due to the thermal and dynamical effects of young OB stars
and type II supernovae.
We confirm that if the remaining gas is not removed from
the remnant of the cloud-cloud collision in our standard model, 
the binary cluster eventually merges  to form a single cluster
because of efficient dynamical friction between the cluster
and the low-density gas. 
In the models with $b/2r_{\rm cl}$ = 0.25,
the developed binary clusters coalesces into a single cluster 
within 0.2 Gyr, for $V_{\rm r}$ $<$ 27 km s$^{-1}$.

The parameter dependencies are summarized as follows (see Figure 4).
Firstly, there is an optimal range of $V_{\rm r}$ 
($10-50$ km s$^{-1}$) for star cluster
formation. The star cluster formation efficiency 
rapidly drops as  $V_{\rm r}$ becomes smaller than
a threshold value ($\sim$  6 km s$^{-1}$). This occurs 
because of the much weaker compression and less efficient shock
dissipation of the colliding gas.
The star formation efficiency  becomes very small
for models with large $V_{\rm r}$ ($>$ $50$ km s$^{-1}$),
since the merging of two clouds does occur at all in these
models. 
Secondly, the star formation efficiency is likely to be 
higher for models with a smaller impact parameter 
$b/2r_{\rm cl}$ (in particular, for $V_{\rm r}$  $<$ $20$ km s$^{-1}$).
Thirdly, a single cluster rather than a binary cluster is likely to be formed
just after the cloud collision in models with smaller $b/2r_{\rm cl}$.
Finally, regardless of model parameters, the star clusters possess
flattened shapes just after their formation.
Our derived higher star formation efficiency can be responsible for
the bound cluster formation after gas removal (e.g., Geyer \& Burkert 2001).

The present study suggests that 
the formation of star clusters in colliding gas clouds with $V_{\rm r}$ $>$ 60 km s$^{-1}$ 
is much less likely for the adopted parameter range of $b/2r_{\rm cl}$. 
As is shown in Figure 5, the two colliding clouds with $b/2r_{\rm cl}$ = 0.25 and $V_{\rm r}$ = 67 km s$^{-1}$
do not show any star-formation in their central regions for 43 Myr evolution.
This is essentially a result of the oblique collision of two clouds with larger $V_{\rm r}$ 
($>$ 60 km s$^{-1}$), they simply graze each other and soon become well separated
without forming strongly shocked and compressed gaseous regions
conducive to star formation. 
Therefore it appears that the formation of a thick gaseous 
layer (with $M_{\rm J}$ of $10^5-10^6$ $M_{\odot}$)  
in colliding clouds with very large $V_{\rm r}$  as proposed by
Kumai et al. 1993 is unlikely.
Our results imply that the formation of star clusters in
colliding clouds is more complex than suggested by the 
previous analytical works of Kumai et al. (1993) and 
Fujimoto \& Kumai (1997).

\section{Discussions}

\subsection{Comparison with previous works}

The dependencies of star cluster formation efficiency 
on $V_{\rm r}$ (or $V_{\rm r}/c_{\rm s}$) 
strongly suggest that  the original proposal by Kumai et al. (1993) 
on collisional cluster formation
should be significantly modified.
Kumai et al. (1993) adopted the following two assumptions to investigate 
the formation processes of star clusters in the LMC/SMC system:
(1) The relative velocity of two interacting clouds
is likely to be more than 50-100 km s$^{-1}$ and
(2) Colliding two gas clouds with such a large relative velocity
can form a compressed (shocked) gas layer, where star formation
can proceed.
Based on these two assumptions,
they claimed that the observed  differences in cluster formation
efficiency between the Galaxy and the LMC/SMC system is
due to large-scale random motions with velocities in the  
Magellanic Clouds, which is not seen in the present-day Galaxy.

We find that the most likely relative velocities
($V_{\rm r}$) of colliding clouds in the interacting LMC/SMC system is 
not 50-100 km s$^{-1}$  but rather  $25-40$ km s$^{-1}$ for a reasonable set of
parameters for the gas clouds.
We have also found that only a minor fraction of colliding gas clouds have
$V_{\rm r}$  more than 100 km s$^{-1}$.
The principal reason for these lower velocities than
those expected by Kumai et al. (1993) 
is that the mass ratio of the SMC to the LMC is 
small enough ($\sim$ 0.1) that
the tidal interaction between the two can not strongly 
disturb the LMC gas clouds.  
These results imply that 
(1) the above first assumption adopted by Kumai et al. (1993)
is invalid and (2) if star clusters in LMC/SMC are formed
from colliding clouds with $V_{\rm r}$ more than 50-100 km s$^{-1}$,
these clusters will be a minority among the young star clusters.

Our simulations have showed that if $V_{\rm r}$ $>$ 60 km s$^{-1}$,
the star formation efficiency in colliding clouds becomes very small
(therefore any bound star clusters are much less likely to be formed).
This is because models with $V_{\rm r}$ $>$ 60 km s$^{-1}$,
result in colliding clouds which just graze with each other and
then become
well separated without forming a strongly shocked gas layer. 
Unless the collision of such gas clouds is close to a head-on,
the clouds cannot form a bound star cluster.
The optimal range of $V_{\rm r}$ 
in our  simulations  suggests that 
the  formation of young star clusters in the LMC/SMC system (but
not in the Galaxy)
are a result the enhanced rates
of cloud-cloud collisions with more moderate relative velocities
($V_{\rm r}$ $\sim$ $10-50$ km s$^{-1}$).
We here stress that this optimal value is true for the clouds
with the adopted size-mass relation Larson (1981), masses
and sizes in the present study.

Kumai et al. (1993) and Fujimoto \& Kumai (1997) found that if
gas clouds make a head-on
collision with $V_{\rm r}$ of $\sim$ 100 km s$^{-1}$,
the Jeans mass ($M_{\rm J}$) of the compressed thin gas layer 
becomes about $10^5-10^6$ $M_{\odot}$, depending on the sound velocity of the gas layer.
The origin of their derived $M_{\rm J}$ for star cluster formation
is their larger adopted value of $V_{\rm r}$. 
The present study has  demonstrated that star clusters are less likely to be formed
in colliding clouds with $V_{\rm r}$ $\sim$ 100 km s$^{-1}$ (also
$M_{\rm J} << 10^5-10^6$ $M_{\odot}$). 
Therefore, a single star cluster with mass 
of $10^5-10^6$ $M_{\odot}$ is less likely to form from gravitationally unstable gas 
with $M_{\rm J}$ of $10^5-10^6$ $M_{\odot}$  in colliding clouds with 
$V_{\rm r}$ $\sim$ 100 km s$^{-1}$.

Our simulations show that star formation starts from the very center
of colliding clouds, where $M_{\rm J}$ is estimated to be $10^3$ $M_{\odot}$
for our adopted isothermal equation of state.    
Stars can continue to  form in the high-density gas at the cloud  center
(with $M_{\rm J}$ of $10^3$ $M_{\odot}$) such that 
the central region of each colliding cloud may be regarded as
``a cluster of sub-clusters'' with the masses of $10^3$ $M_{\odot}$. 
These ``sub-clusters'' are all located in the very center of the colliding cloud,
and consequently may quickly merge with one another to form a single
massive cluster. Thus, massive clusters ($10^5-10^6$
$M_{\odot}$) may result from the 
merging of numerous small sub-clusters in the cloud centers since
$M_{\rm J}$ of the gas is less than $10^5-10^6$  $M_{\odot}$.

Since the resolution of the present simulation is at most $\sim$ $10^2$ $M_{\odot}$
in mass and $\sim$  1 pc in size for models with $M_{\rm cl}$ = $10^6$ $M_{\odot}$,  
we can not investigate the details of the merging process of
these sub-clusters.  It is necessary to represent such a small
cluster not as a single stellar particle, but as a collection of 
stellar particles (with the total number of $10^2-10^3$),
to allow a rigorous investigation of the structural and 
kinematical properties of the 
remnants of multiple mergers between numerous small clusters. 
We leave to a future numerical study, with the total particle 
numbers of more than $10^6$,
to confirm (1) whether such small clusters are first formed in
the central regions of colliding clouds and (2) how these 
clusters can merge with one another to form a single
massive cluster.

\subsection{Origin of the observed mass-size relation of star clusters}

We have found  that the star formation efficiency in colliding gas clouds 
becomes  smaller in the model with larger $V_{\rm r}$ (or $V_{\rm r}/c_{\rm s}$) 
for $V_{\rm r}$ $>$ $30$ km s$^{-1}$.
This result provides a new clue to
the origin of the observed mass-size relation of star and globular
clusters. 
There is observational evidence suggesting only a  weak correlation 
between the mass ($M_{\rm st}$) and size ($R_{\rm st}$) of
young star clusters and globular clusters
(e.g., $R_{\rm st} \propto {M_{\rm st}}^{0.1\pm 0.1}$ for young clusters;
Zepf et al. 1999). 
Ashman \& Zepf (2001) pointed out that if star clusters and globular clusters
are formed from molecular gas clouds with the observed size-mass
relation of $r_{\rm cl} \propto {M_{\rm cl}}^{0.5}$ (Larson 1981),
then the star formation efficiency should be lower in  smaller gas clouds
to  reproduce  the observed mass-radius relation. 
The sound velocity and gas  temperature are smaller in
smaller self-gravitating gas clouds (Larson 1981).
However,  $V_{\rm r}$ is controlled by global  galactic dynamics
and therefore is independent of gas cloud mass.
Our results suggest that star formation efficiency is
lower for smaller gas clouds due to the larger $V_{\rm r}/c_{\rm s}$
(for $V_{\rm r}$ $>$ 30 km s$^{-1}$).
We note that the origin of the observed scaling relation of star clusters
could  be closely associated with the star formation efficiency
dependent on $V_{\rm r}/c_{\rm s}$ in  colliding clouds.

Following  the simple analytic argument by Ashman \& Zepf (2001),
we can discuss this point in a more quantitative manner.
First we define $\epsilon$ to be  the star formation efficiency
in colliding gas clouds;
\begin{equation}
\epsilon = \frac{M_{\rm st}}{M_{\rm cl}}.
\end{equation}
We then assume that the size of a cluster depends upon the star formation
efficiency of the cluster's progenitor cloud such that 
(e.g., Hills 1980);
\begin{equation}
\frac{R_{\rm st}}{r_{\rm cl}} \simeq  {\epsilon}^{-1} .
\end{equation}
Based on our simulations, we can write the dependence of $\epsilon$ on
$V_{\rm r}/c_{\rm s}$ as follows; 
\begin{equation}
\epsilon \propto (\frac{V_{\rm r}}{c_{\rm s}})^{\alpha}. 
\end{equation}
Therefore, the dependence of $R_{\rm st}$ on $M_{\rm cl}$  is: 
\begin{equation}
R_{\rm st} \propto r_{\rm cl} {\epsilon}^{-1} \propto {M_{\rm cl}}^{1/2+\alpha /4}
\propto {M_{\rm st}}^{\frac{2+\alpha}{4-\alpha}}
\end{equation}
Here we assume that (1) $r_{\rm cl} \propto {M_{\rm cl}}^{0.5}$ (Larson 1981),
(2) $c_{\rm s} \propto {M_{\rm cl}}^{0.25}$ predicted from the virial theorem
and Larson's relation above, 
and (3) $V_{\rm r}$ does not depend on initial cloud mass.
It is clear from the above equation that $\alpha$ should be
$\sim$ $-2$  to explain the apparent lack of a mass-radius 
relation in young clusters (Zepf et al. 1999).
Although our simulations suggest that  $\alpha$ takes a negative value,
they do not enable us to derive a robust value (or range) for
$\alpha$ because of our relatively small parameter space. 
We plan to estimate  $\alpha$ in more sophisticated future simulations
with a much wider parameter space of colliding gas clouds.

\subsection{Formation of highly flattened star clusters in LMC/SMC}

Origin of the flattening of LMC star clusters
have been discussed in several authors (e.g., Frenk \& Fall 1982; 
Kontizas et al. 1989).
de Oliveira et al. (2000) investigated star cluster encounters
in their {\it purely collisionless} simulations, and found that 
binary clusters can merge with each other to form a single
cluster with higher ellipticity.
They also showed that if the mass ratio of two merging clusters
with orbital eccentricities of $0.6-0.9$ (i.e., a bound orbit)
is close to 0.1 (``minor merging''), the merger remnant shows
an ellipticity consistent with observations of LMC clusters.
The present study has demonstrated that binary clusters can be formed
from cloud-cloud collisions.
Our study, and that of de Oliveira et al. (2000), suggest that the
observed globular and populous clusters with higher  ellipticity
in the Magellanic Clouds may originate from the merging of binary
clusters formed from cloud-cloud collisions.
These two studies also imply that the difference in the shapes of clusters
between the Galactic halo/disk globular clusters and the young
Magellanic Cloud clusters may be due to the fact that only 
LMC/SMC clusters have experienced the past merging
of star clusters. 

It is, however, unclear whether binary clusters with the mass
ratio of $\sim$ 0.1 can actually be formed in colliding gas
clouds.
All of our models involve ``major mergers''
of two clouds such that the incipient star clusters in the centers of the two
clouds have similar masses. 
Also, we have only investigated the collision of two gas clouds
with identical radial density profiles. This does not allow us to
predict the final mass ratio of two clusters formed from unequal-density 
mergers of gas clouds.
We need to investigate extensively a set of numerical simulations
with a wider range of parameters such as 
the mass ratio and the density profiles of colliding two clouds
in order to confirm whether a binary cluster formed from a 
cloud-cloud collision has the mass ratio of $sim$ 0.1.
Our future more sophisticated simulations including chemical evolution,
magnetic fields,  dynamical evolution of hierarchical/fractal  
structures within a cloud, and feedback effects from massive 
stars and supernovae will address the origin of flattened shapes
of LMC/SMC clusters in a more quantitative way.

\placefigure{fig-3}
\placefigure{fig-4}

\section{Conclusions}

We have used two different sets of numerical simulations to
understand the origin of the physical properties of star clusters
in the Magellanic Clouds. Although the present model of star cluster
formation in colliding two clouds is simplified in a number of areas,
we have revealed some essential aspects of star cluster formation
in colliding gas clouds. 

The main conclusions  are summarized as follows.
 
(1) An oblique collision between two identical clouds can be significantly
enhanced during the tidal interaction between the LMC and the SMC.
Cloud-cloud collisions with radial velocities ($V_{\rm r}$) of
$25-40$ km s$^{-1}$ are most common in our models. 
These results imply that the origin of young star clusters in the LMC/SMC 
systems may stem from the enhanced collision rates of gas clouds with
moderately large relative velocities.
Our results also suggest that the original proposal by Kumai et
al. (1993) on collisional cluster formation should be modified.

(2) Compact, bound star clusters can be formed in the centers of colliding
gas clouds as a result of strong gas shocks, compression, and dissipation
during the collision. The initial impact parameter of two colliding clouds
determines whether the star clusters form a single cluster, a 
binary cluster, or two isolated star clusters.
For example, for a smaller impact parameter, the incipient clusters
soon merge with each other to form a single, more massive cluster.

(3) Star formation efficiency in colliding clouds depends on 
the initial ratio of the relative velocity of the clouds to the sound
speed of the gas ($V_{\rm r}/c_{\rm s}$). This dependency is in
the sense that the star formation efficiency is lower for
models with larger $V_{\rm r}/c_{\rm s}$.
The derived dependence on $V_{\rm r}/c_{\rm s}$ provides a new clue
to the origin of the observed mass-size relation of young star clusters.

\acknowledgments
We are  grateful to the anonymous referee  for valuable comments,
which contribute to improve the present paper.
K.B. and W.J.C. acknowledge the Large Australian Research Council (ARC).
MB acknowledges assistance from the Swinburne Research and 
Development Grant System.

\newpage


\clearpage


\figcaption{
{\it Left}: Mass distributions of interacting two galaxies at $T$ = 0.61 Gyr (upper)
and 0.92 Gyr (lower). 
Here $T$ represents the time that has elapsed since the simulation starts
(i.e., the two disks begin to interact with each other).
Only stellar and gaseous components are plotted in these panels.
Scales are given in  units of  the LMC's disk size, 
and so each  frame measures 30 kpc on a side.
The center of the small circle  represents the  position of the 
 SMC and the circle size is equal to the disk size of the SMC. 
{\it Right}: 
Time evolution of cloud-cloud collision rate in the interacting galaxies (top),
the number distribution of the relative velocity of colliding two clouds  (middle),
and that  of the impact parameter of the two colliding  clouds (bottom) for
the interacting galaxies at $T$ = 0.92 Gyr.
The impact parameter ($b$) is given in units of the diameter of a cloud (i.e., $2r_{\rm cl}$). 
\label{fig-1}}

\figcaption{
Distribution of gas (cyan) and new stars formed from gas (magenta)
of colliding two clouds in the standard model with
$b/2r_{\rm cl}$ = 0.25 and $V_{\rm r}$ = 20 km s$^{-1}$
projected onto the $x$-$y$ plane, at each time indicated in
the upper left corner of each panel. One frame measures 583 pc on a side. 
\label{fig-2}}

\figcaption{
Time evolution of the star formation rate of the standard model
($b/2r_{\rm cl}$ = 0.25 and $V_{\rm r}$ = 20 km s$^{-1}$).
\label{fig-3}}

\figcaption{
Dependence of the mass fraction of new stars within two colliding  clouds  on
the relative velocity of the clouds for a given  impact parameter ($b$).
The mass fraction here is defined as $M_{\rm s}/M_{\rm g}$, where
$M_{\rm s}$ and $M_{\rm g}$ are the total mass of new stars
formed before $T$ = 56 Myr and initial gas mass of the clouds, respectively.
The results are shown for $b/2r_{\rm cl}$ = 0.1 (long dash),
 0.25 (short dash), 0.5 (dotted), and 0.75 (solid).
Note that there is an optimal range ($10-50$ km s$^{-1}$) for the efficient
star (or star cluster) formation in colliding clouds.
\label{fig-4}}

\figcaption{
The same as Figure 2 but for the model with 
$b/2r_{\rm cl}$ = 0.25 and $V_{\rm r}$ = 67 km s$^{-1}$.
Note that no star formation occurs in this large $V_{\rm r}$ model during the 
collision of two gas clouds.
\label{fig-5}}

\newpage
\plotone{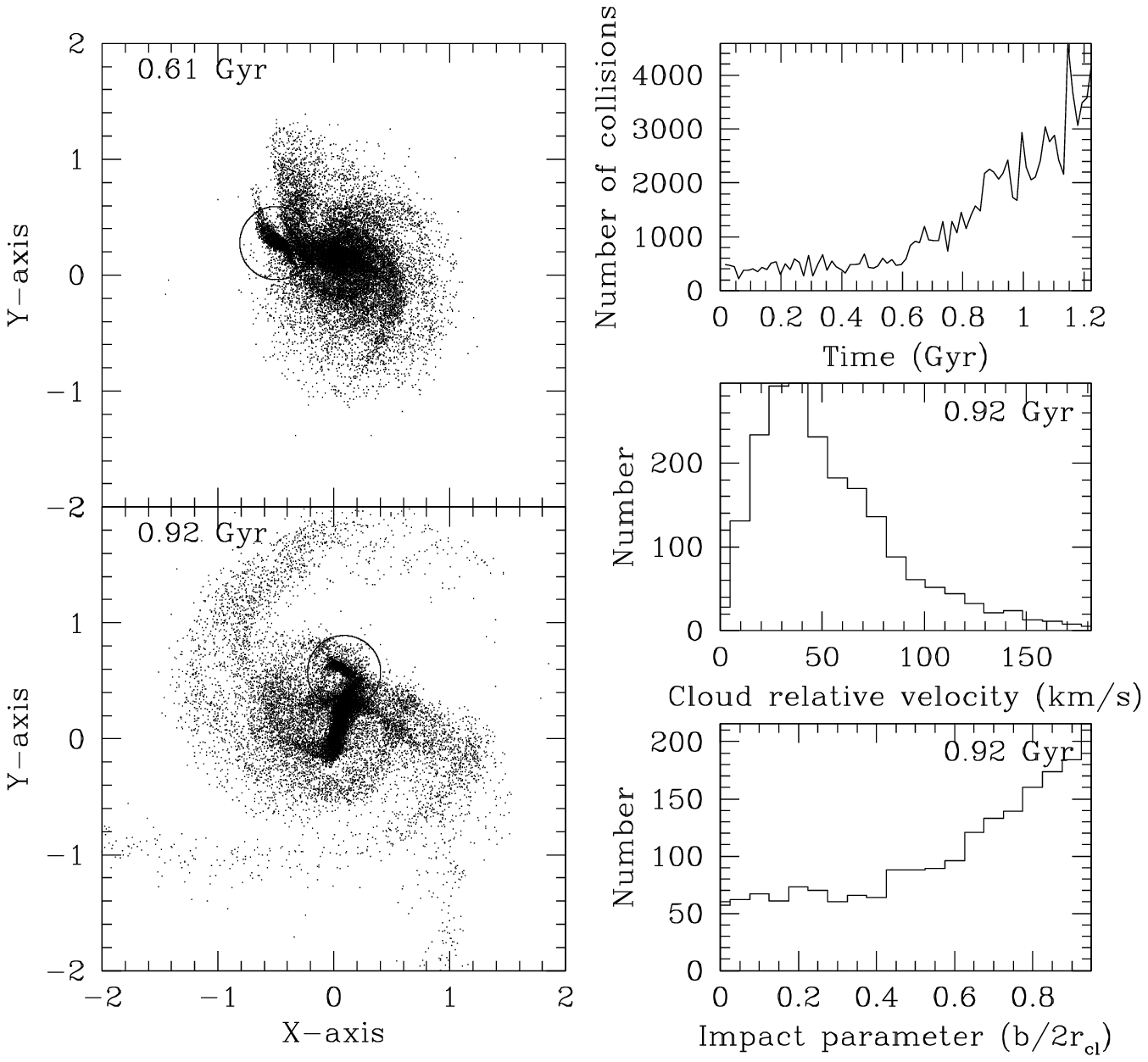}
\newpage
\plotone{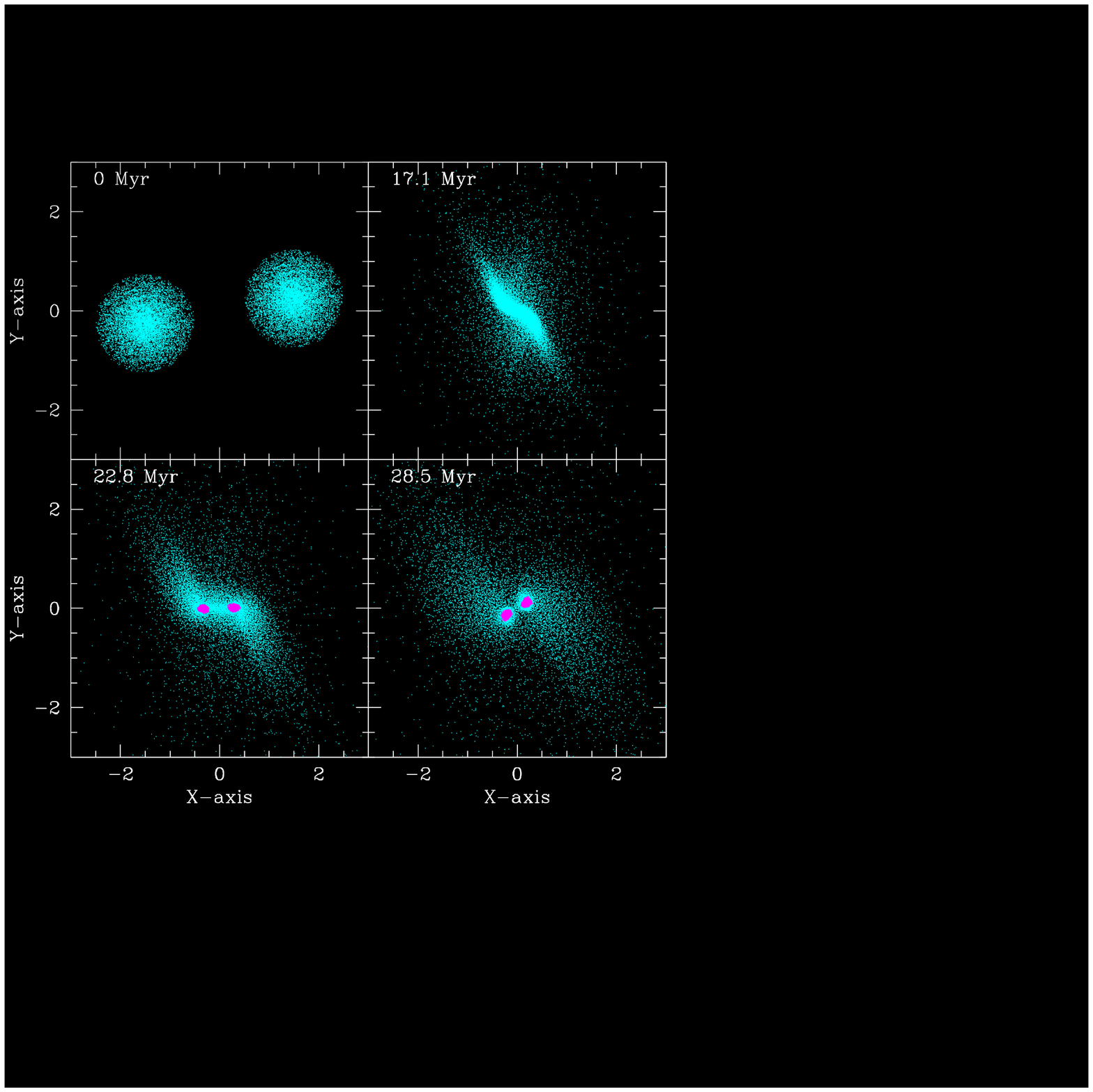}
\newpage
\plotone{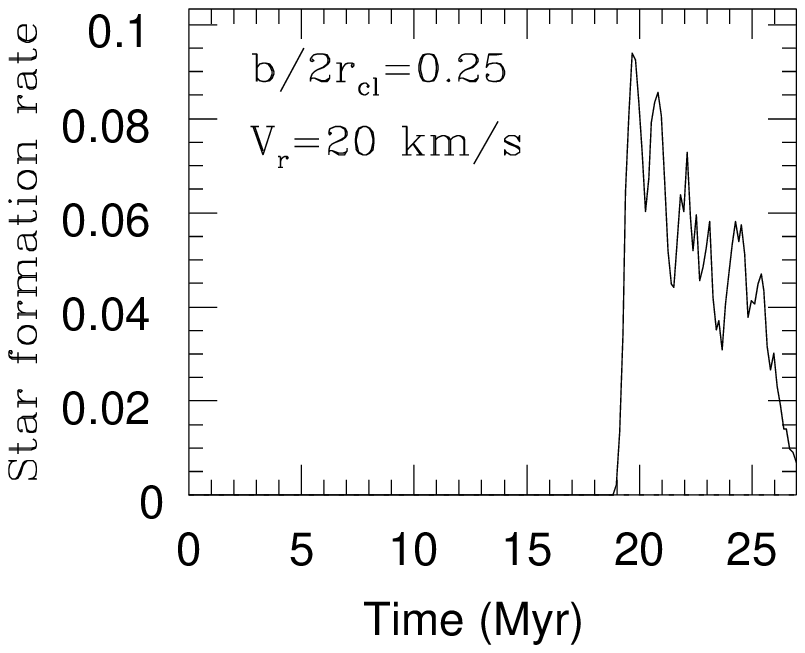}
\newpage
\plotone{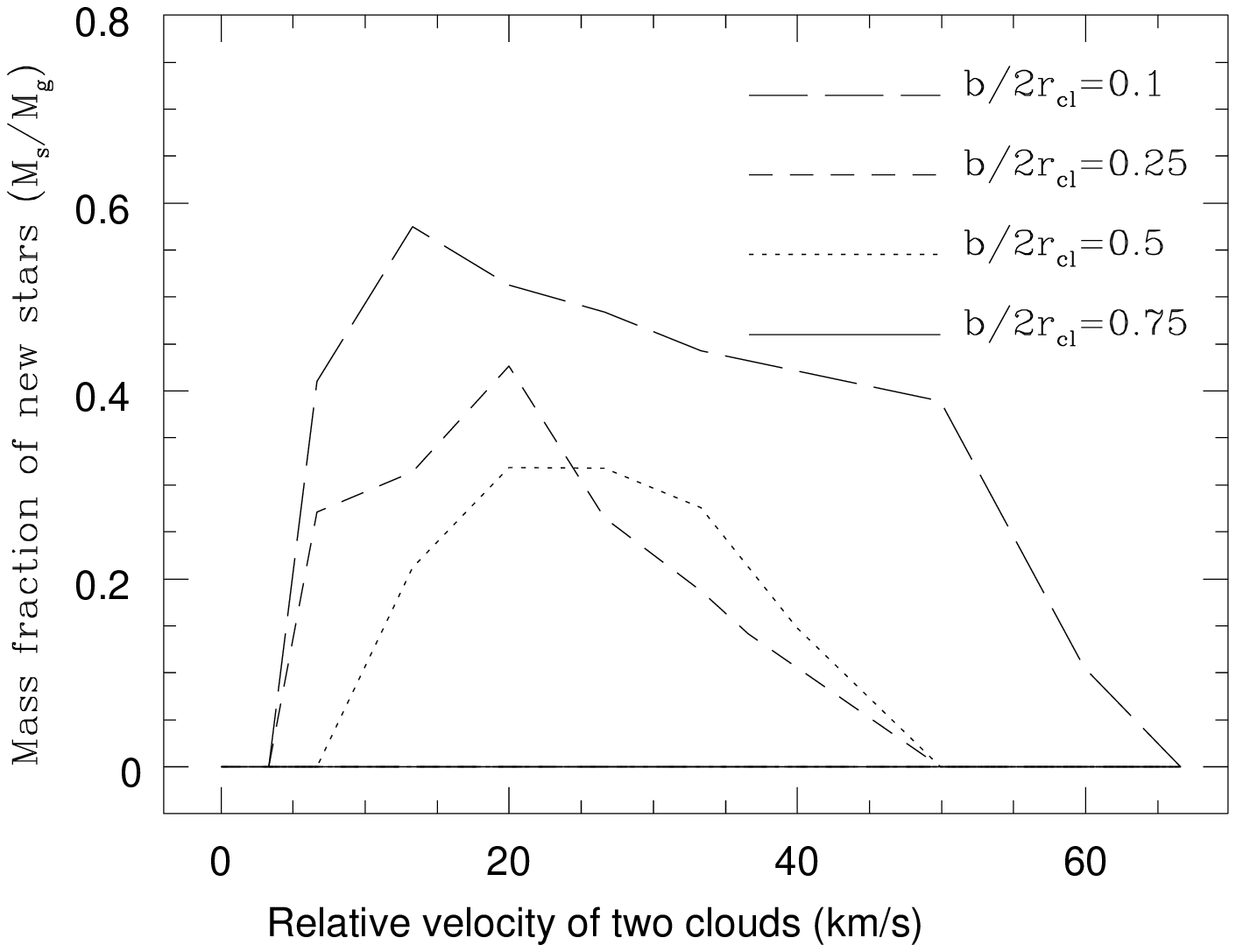}
\newpage
\plotone{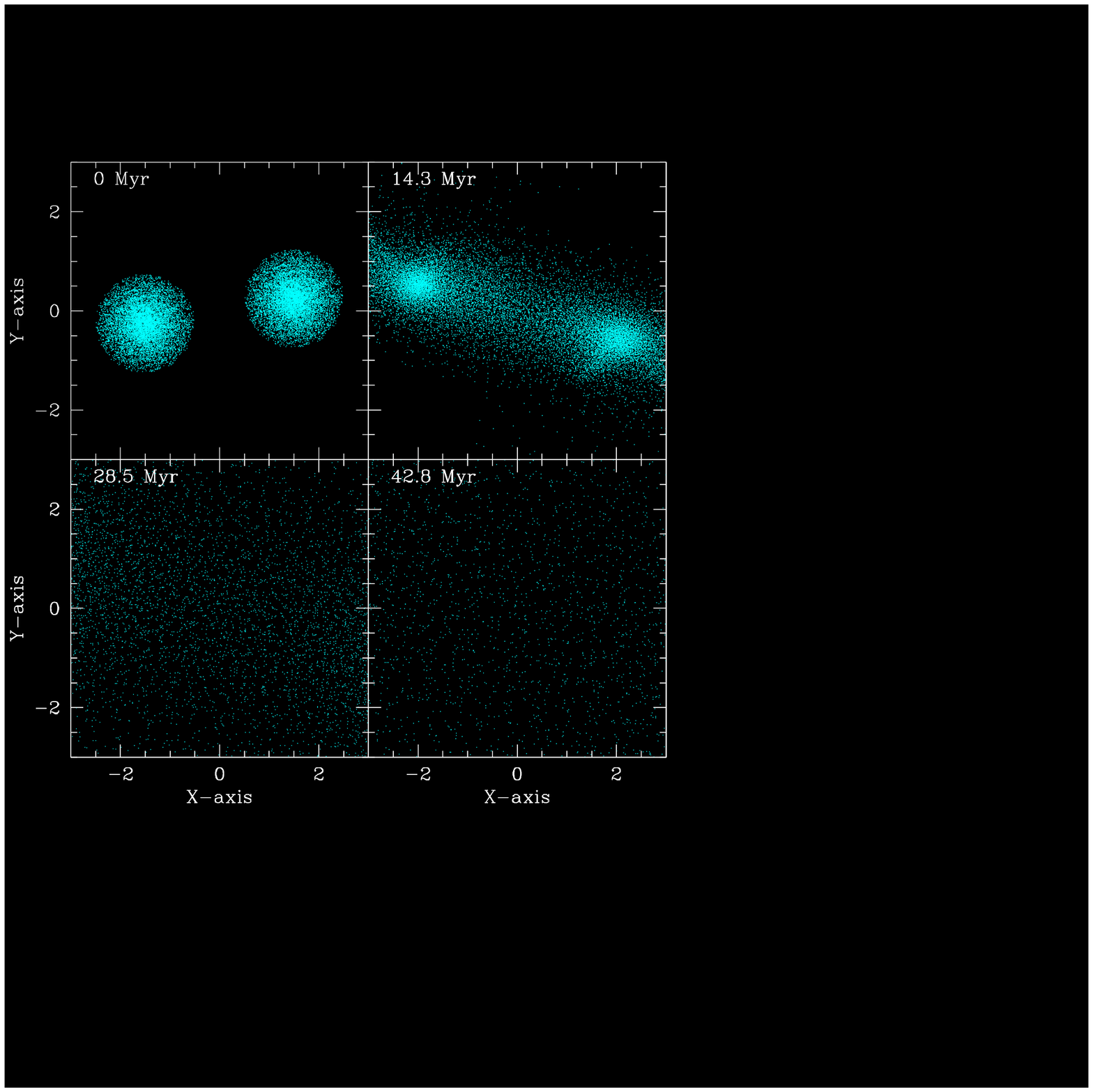}

\end{document}